\documentclass{article}

\usepackage{xcolor} 
\usepackage{natbib}
\usepackage{graphicx}
\usepackage{amsmath, amsfonts, amssymb, mathrsfs, rotating, setspace}
\usepackage[margin=2.5cm]{geometry}
\usepackage{url} 
\usepackage{tikz}
\usepackage{rotating}
\usepackage{adjustbox}

\usetikzlibrary{calc}
\usetikzlibrary{positioning}
\usetikzlibrary{arrows}

\newcommand\independent{\protect\mathpalette{\protect\independenT}{\perp}}
\def\independenT#1#2{\mathrel{\rlap{$#1#2$}\mkern2mu{#1#2}}}

\title{Propensity weighting plus adjustment in proportional hazards model is not doubly robust}

\author
{Erin E Gabriel$^{1,*}$, 
Michael C Sachs$^1$, 
Ingeborg Waernbaum$^2$ \and Els Goetghebeur$^3$, Paul F Blanche$^1$, Stijn Vansteelandt$^3$ \and  Arvid Sjölander$^4$, and Thomas Scheike$^1$\\
$^{1}$Section of Biostatistics, Department of Public Health, \\ University of Copenhagen, Denmark  \\
$^{2}$Department of Statistics, Uppsala University, Sweden \\
$^3$Department of Applied Mathematics, Computer Science and Statistics, \\ Ghent University, Belgium \\
$^4$Department of Medical Epidemiology and Biostatistics, \\ Karolinska Institutet, Sweden
}
\date{}

\begin{document}



\maketitle

\noindent\textbf{Abstract:}
Recently, it has become common for applied works to combine commonly used survival analysis modeling methods, such as the multivariable Cox model and propensity score weighting, with the intention of forming a doubly robust estimator of an exposure effect hazard ratio that is unbiased in large samples when either the Cox model or the propensity score model is correctly specified. This combination does not, in general, produce a doubly robust estimator, even after regression standardization, when there is truly a causal effect. We demonstrate via simulation this lack of double robustness for the semiparametric Cox model, the Weibull proportional hazards model, and a simple proportional hazards flexible parametric model, with both the latter models fit via maximum likelihood. We provide a novel proof that the combination of propensity score weighting and a proportional hazards survival model, fit either via full or partial likelihood, is consistent under the null of no causal effect of the exposure on the outcome under particular censoring mechanisms if either the propensity score or the outcome model is correctly specified and contains all confounders. Given our results suggesting that double robustness only exists under the null, we outline two simple alternative estimators that are doubly robust for the survival difference at a given time point (in the above sense), provided the censoring mechanism can be correctly modeled, and one doubly robust method of estimation for the full survival curve. We provide R code to use these estimators for estimation and inference in the supporting information. 

\clearpage
\section{Introduction} 
Many concepts from causal inference have now made their way into practice in applied epidemiological and medical research. One such concept is doubly robust estimation. In this paper, we will focus on the concept of double robustness with regard to accounting for confounding rather than missing data or censoring. There is a large literature on such estimators, and this is the basis for many well-known causal estimation and inference methods, such as double machine learning and TMLE. Doubly robust (DR) estimators generally use some form of working propensity score model, i.e. a model for the exposure, in addition to a working outcome model that includes some adjustment for confounders. These two working models are combined to provide a consistent estimator of the exposure effect on the outcome if either the propensity model or the outcome model is correctly specified and sufficiently accounts for all confounding. Our goal is not to outline or review all existing DR estimators or methods or derive new such estimators, but rather to point out some seemingly common misconceptions about what makes an estimator doubly robust in the setting of survival data. 

We believe that \citet{robins1992estimating}, in their Section 4, were the first to propose a doubly robust estimator for confounding. Although they did not call it doubly robust, they did note the property. As outlined in \citet{seaman2018introduction}, \citet{scharfstein1999adjusting} also noted that an estimator proposed in \citet{robins1994estimation} was consistent if either the missingness model or the imputation model used in it was correctly specified not requiring that both were simultaneously correct. This property became known as “double robustness” \citep{robins2000robust} both for missingness and confounding. 

Double robustness is a very appealing concept, and concerns over potential model misspecification, and thus residual confounding, have led many applied researchers to want to use doubly robust methods. However, potentially a lack of user-friendly software or a misunderstanding of what makes an estimator doubly robust have led some researchers to use propensity score weighted regression adjusted estimators, such as a weighted partial maximum likelihood fit of a Cox model, possibly to gain robustness. Applied papers using such a technique \citep{ionescu2021association, estruch2018primary, simon2020association, sagara2016patient, vaughan2015application}, generally do not provide a reasoning for the method selected. Exceptionally,  \citet{simon2020association} directly cite double robustness as the reason for using inverse probability of treatment weights (IPTW) based on the propensity score in combination with an adjusted Cox model from which they report adjusted hazard ratios. Unfortunately, this combination is not generally doubly robust for the causal conditional hazard ratio when there is a causal effect of the exposure. However, it does provide robustness against model misspecification under the null for some types of censoring, as we will show below. 

The main point of this work is to show clearly that it is not simply the combination of propensity score and an adjusted survival outcome model that is required for an estimator to have robustness against misspecification of either model and to caution researchers against belief in such ad hoc estimator constructions. We do this by providing clear definitions for all terms involved in the discussion. We then specifically demonstrate, via simulations, that the combination of propensity score weighting and a proportional hazards model, fit either via maximum likelihood or partial likelihood, will not generally produce a doubly robust estimator. Additionally, we demonstrate alternatives for estimators for the difference in survival probabilities at a fixed time point that are doubly robust. We note that these are not new methods developed herein but rather a review of some easy-to-use alternatives that provide the double robustness so many applied articles seem to be seeking and failing to obtain with their ad hoc estimators. It is not our intention to comment directly about any one paper or result, even in the above-cited works, as there are many more examples in the literature. Rather, we commend researchers for wanting to use more robust methods and empathize with statistical practitioners about the lack of user-friendly software for more complex DR methods. 

After defining notation in Section \ref{not}, we give definitions of the terms we will use throughout and review related work in the setting of randomized controlled trials. In Section \ref{coxsec}, we detail why combining IPTW and an adjusted Cox model fit either via partial likelihood or full likelihood via a parametric proportional hazard survival model is not generally doubly robust. In Section \ref{rec} we provide some recommendations for methods that are doubly robust under completely independent censoring or when the censoring mechanism can be modeled correctly. In Section \ref{sims}, we provide simulations to demonstrate our claims, and we demonstrate the use of our suggested estimators in Section \ref{Real} in a real data example. Finally, in Section \ref{summary}, we summarize our main points and in Section \ref{dis} provide further discussion of several points related to our main topic.

\section{Notation and preliminaries} \label{not}
Let $T_i$ be time until the event of interest for subject $i$, where $i = 1, \ldots, n$. We will occasionally omit the index where not needed. When there is (right) censoring, we will observe $\widetilde{T}_i =\mbox{min}(T_i, C_i)$ and $\Delta_i = I_{T_i\leq C_i}$ where $C_i$ is the censoring time. Let $X_i$ be the exposure of interest, which we assume is binary throughout. Let $T_i(x)$ be the counterfactual or potential value of $T_i$ for subject $i$ had $X_i$, potentially counter to fact, been set to $x$. We will throughout make the consistency assumption that $T_i(x) = T_i$ for individuals with $X_i=x$. Further, let $D = \{l:\Delta_l = 1\}$ i.e., the index set of subjects with uncensored events times, and for any time $k \geq 0$ let the risk set be $R(k) = \{j:\Tilde{T}_j \geq k\}$. Let $\boldsymbol{Z}_i=\{Z_{i1}, \ldots, Z_{im}\}$ be a set of measured baseline covariates.

\subsection{Terms and concepts}
\paragraph{Confounding}
When fitting an model for the outcome only, to estimate the causal effect of interest, one must, in general, correctly specify the relationships between the outcome and a sufficient set of confounders and the exposure to control confounding and obtain a consistent estimator; please see \citet{witte2019covariate} for more details on what constitutes a sufficient adjustment set. 
The covariates $\boldsymbol{Z}$ being sufficient for confounding control means
\begin{eqnarray}
\label{confounding}
T(x) \independent X|\boldsymbol{Z}.
\end{eqnarray}
All confounders are measured sufficiently to allow inclusion in an outcome or propensity model using some functional form only depending on observed data. Thus, we assume there are no unmeasured confounders. 

\paragraph{Censoring}
In survival settings we have the additional concern of censoring, as the time we observe is generally the minimum of the censoring time and the event time. When the censoring time is covariate-dependent, a similar concept can be considered for censoring as for confounding, if there is a sufficient set of covariates $\boldsymbol{Z}$ such that 
\begin{eqnarray}
\label{eq:cen}
T \independent C|X,\boldsymbol{Z}.
\end{eqnarray}

\paragraph{Correctly specified}
A \textbf{correctly specified} model has the correct functional form to model the relationship between the outcome and included covariates. Specifically, we will use the terms `misspecified model', `misspecification', and `correct model' throughout. We will be specific here to avoid any confusion. Let the distribution of the true data-generating mechanism of a generic outcome $Y$ given $X$ and $\boldsymbol{Z}$ be $f(y;X,\boldsymbol{Z}).$ A regression model for $Y$, say $g(X,\boldsymbol{Z};\beta)$, then corresponds to some set of probability distributions, $\Omega$, and is correctly specified if it contains the true distribution, i.e., $f(y; X,\boldsymbol{Z}) \in \Omega$.
Thus, multiple models may be simultaneously correctly specified in some cases, in particular in regression settings that only specify the mean model. However, this is not usually the case in survival, where we most often specify a model for the hazard. 

In the special case that an outcome model contains 
the sufficient set $\boldsymbol{Z}$ to make independence relation \eqref{confounding} hold, in addition to $X$, and is correctly specified we say that this model is \textbf{correctly specified for confounding}. A propensity score model for $X$ can also be, given our definition, correctly specified for confounding if it includes the sufficient set $\boldsymbol{Z}$ and is correctly specified. 

Similarly, if a correctly specified outcome model contains a sufficient set $\boldsymbol{Z}$ such that independence relation \eqref{eq:cen} holds, we will call this \textbf{correctly specified for censoring}. A model for $C$ can also be correctly specified for censoring if it includes the sufficient set $\boldsymbol{Z}$ and is correctly specified. 

Any method of fitting a survival outcome model that is both correctly specified for censoring and correctly specified for confounding will provide consistent estimates of some causal parameter. A model, however, may be correctly specified for censoring or correctly specified for confounding, neither, or both, and thus, we consider these scenarios.

We note that if a Cox model is correctly specified for confounding, only under very specific distributional and independence assumptions can another Cox model be correctly specified, for confounding or even only statistically, and not be the same model in the same data. For example, if there is a purely prognostic variable that has the positive stable distribution, then two correctly specified for confounding Cox models that are not the same may exist \citep{henderson1999effect}. However, having noted this we move forward under the simplifying assumption that there is only one correctly specified Cox model. Thus, we will refer to `the true' causal estimand.

\paragraph{Collapsible}
Although collapsibility is a wider topic, we focus on the collapsibility of the hazard ratio as we are only considering proportional hazard models. The hazard ratio is not collapsible in the sense that the marginal hazard ratio, marginalizing over, for example, $\boldsymbol{Z}$, does not generally equal the conditional hazard ratio given $\boldsymbol{Z}$. This is true even if the conditional hazard ratio is constant over levels of $\boldsymbol{Z}$, $\boldsymbol{Z}$ is independent of $X$, and both the marginal and conditional hazard ratios are constant in time. For clarity, the marginal hazard ratio is the hazard ratio for a single variable unconditional on other
variables, e.g. a correctly specified Cox model containing only the exposure will estimate the marginal hazard
ratio for the exposure. A more in-depth discussion of the lack of collapsibility for hazard ratios can be found in \cite{sjolander2016note}, while a more general treatment can be found in \cite{neuhaus1993geometric} for collapsibility, and in \cite{daniel2021making} for collapsibility and marginal hazard ratios. 


\paragraph{Regression standardization} Regression standardization, also called regression estimators, regression imputation estimators, and the g-formula estimators \citep{tan2010bounded}, is the averaging over the observed or known distribution of the covariates included in a regression model that are not the exposure while holding the exposure constant. This allows for estimating causal marginal effects from models with interactions between the exposure and the covariates and from noncollapsible estimand/models, including covariates. For example, standardization can be used to obtain estimates of the causal marginal survival difference at time $t$ from a Cox model. The standardization estimator takes the form $\widehat{p\{T(x)>t\}}=\widehat{E}\{\widehat{p}(T>t|X=x,\boldsymbol{Z})\}=\frac{1}{n} \sum_{i=1}^n \widehat{S}(t|x,\boldsymbol{Z}_i;\boldsymbol{\widehat{\gamma}}),$
where $\widehat{S}(t|x,\boldsymbol{Z}_i;\boldsymbol{\widehat{\gamma}})$ is the estimated survival probability at time $t$ corresponding to the parametric or semi-parametric model that was fit. 


\paragraph{Inverse probability of treatment weighting}
Consider that we have a propensity score for receiving exposure given by
\begin{eqnarray}
p(X=1|\boldsymbol{Z})=g(\boldsymbol{Z};\boldsymbol{{\alpha}}), \label{ipw}
\end{eqnarray}
for some known link function $g(\cdot)$. Using this model, the IPTW are given by 
\begin{eqnarray}
W(X,\boldsymbol{Z};\widehat{\boldsymbol{\alpha}})=\frac{X}{g(\boldsymbol{Z};\widehat{\boldsymbol{\alpha}})}+\frac{1-X}{1-g(\boldsymbol{Z};\widehat{\boldsymbol{\alpha}})}, \label{weights}
\end{eqnarray}
where $\boldsymbol{\widehat{\alpha}}$ is some consistent estimator of the parameter vector $\boldsymbol{\alpha}$ when the propensity score model is correctly specified for confounding, e.g. the solution to a set of MLE score equations. 

The propensity score can be used in various ways including matching to create a new unconfounded data set and weighting of the score or estimating equations. Weighting is the method we focus on in this work. Heuristically, IPT weighting is targeting the same estimand that would be obtained in a randomized trial. Thus, no method using IPT weighting of the estimating equations for an observational study can be expected to decrease bias over applying the unweighted estimating equations in a perfectly randomized trial.

\paragraph{Estimands}
An estimand is the target of estimation; a causal estimand is the target of causal estimation and is almost always stated in terms of counterfactuals. As the majority of applied papers weighting the partial likelihood Cox model or including a propensity score as a covariate in the Cox model report conditional (log) hazard ratios, this will be our primary target of discussion. We will refer to this as the causal conditional (log) hazard ratio. For clarity, throughout, we assume that the true hazard for the failure time $T$ is given by  
$
\lambda^{\star}(t|X,\boldsymbol{Z}; \boldsymbol{\omega})
$
making the causal conditional log hazard ratio, at a given time point $t$ for the true $\boldsymbol{\omega}$, 
$
\psi(t,\boldsymbol{Z};\boldsymbol{\omega})=\log\left\{\frac{\lambda^{\star}(t|1,\boldsymbol{Z}; \boldsymbol{\omega})}{\lambda^{\star}(t|0,\boldsymbol{Z}; \boldsymbol{\omega})} \right\}.
$
This may or may not depend on time. However, in order for any of the proportional hazards working models we will propose to be correctly specified this will need to be independent of time. As pointed out by multiple authors, hazard ratios have a difficult causal interpretation \citep{MartinussenHR20, hernan2010hazards}, particularly when they are time dependent. Thus, when we refer to $\psi(t,\boldsymbol{Z};\boldsymbol{\omega})$ as the causal conditional log hazard ratio, it will be with the understanding that the causal interpretation of this parameter is non-trivial, outside the proportional hazard setting. Additionally we refer to the causal null as: $T \independent X| \boldsymbol{Z}$, (or just the null), which, in terms of hazards, is expressed as $\psi(t,\boldsymbol{Z};\boldsymbol{\omega}) = 0$ for all $t$.

Due to the difficult causal interpretation of conditional hazard ratios, we also target the causal marginal difference of survival probabilities at some time-point $t$: $$\zeta(t)=p\{T(1)>t\}-p\{T(0)>t\};$$ where the null is $\zeta(t)=0$. This will also be the estimand for which we provide alternative doubly robust estimators. We will consider the regression standardization estimators for this estimand following the propensity score weighting of an adjusted PH model. 

\subsection{Proportional hazard models in randomized clinical trials} \label{rand}
There have been a number of works considering misspecified proportional hazards models fit with full or partial likelihood in randomized clinical trials. Most of these works, to our knowledge, focus on hypothesis testing rather than estimation. \citet{kong1997robust}, prove that any working model of the form 
\begin{eqnarray}
\lambda(t|X,\boldsymbol{Z}; \beta, \gamma, \nu) =  \lambda_0(t, \boldsymbol{Z}, \boldsymbol{\gamma}, \nu)\, \exp(\beta X) \label{general}
\end{eqnarray}
and fit either via maximum likelihood, either with full or partial specification of the baseline hazard $\lambda_0(t, \boldsymbol{Z}, \boldsymbol{\gamma}, \nu)$, will be consistent for the causal conditional log hazard ratio when $\psi(t,\boldsymbol{Z}; \boldsymbol{\omega}) = 0$, under specific conditions on censoring. The censoring time 
must have a hazard without an interaction between $X$ and $\boldsymbol{Z}$ with the form 
\begin{eqnarray}
\lambda_C(t|X,\boldsymbol{Z}) = \alpha_0(t) + \alpha_1(t,X) +\alpha_2(t,\boldsymbol{Z}), \label{sludcens}
\end{eqnarray} 
such that $P(C>t|X,\boldsymbol{Z}) = \exp{\{-A_0(t) -A_1(t,X) -A_2(t,\boldsymbol{Z})\}},$ and $\alpha_1(t,0)=A_1(t,0)=0$. We will refer to this as type A covariate-dependent censoring, a special case of which is censoring that only depends on $X$ or $\boldsymbol{Z}$. We reproduce their proof in the simpler settings we consider in supporting information. \citet{kong1997robust} show that under these same assumptions, tests based on the estimated coefficient of $X$ in such models have the correct type I error rate when the proper variance estimator, which they provide, is used. 

A simpler modeling framework was investigated in \citet{dirienzo2001effects}, in which fully proportional Cox models were considered, and the same result was found. Neither \citet{dirienzo2001effects} nor \citet{kong1997robust} consider estimation, as their focus is on testing in a randomized trial for any effect of the treatment and not estimating that effect. For this reason, they also do not consider the consistency of the estimator under the alternative, i.e. any setting where $\psi(t,\boldsymbol{Z};\boldsymbol{\omega}) \neq 0$.

\section{IPTW and simple proportional hazards models} \label{coxsec}
We now consider a working proportional hazards outcome model of the form: 
\begin{eqnarray}
\lambda(t|X,\boldsymbol{Z}; \beta, \boldsymbol{\gamma}, \nu) =  \lambda_0(t; \nu)\, \exp\{\beta X+m(\boldsymbol{Z};\boldsymbol{\gamma})\},  \label{cox}
\end{eqnarray}
where $\lambda_0(t; \nu)$ is the baseline hazard and is either specified according to some parametric model or left unspecified in the case of the Cox model. $m(\boldsymbol{Z};\boldsymbol{\gamma})$ is a known function of an unknown parameter vector $\boldsymbol{\gamma}$. Here, we do not consider interactions between $X$ and $\boldsymbol{Z}$ for clarity, as multiple parameters would need to be considered if interactions were allowed. However, there is nothing restricting the use of interactions in the working models; the covariates considered would simply need to be expanded. We will also use the IPTW given in \eqref{weights} that are based on the working model given above in \eqref{ipw}. All estimators that we consider in this Section are consistent for their estimands, if the outcome model is correctly specified for confounding and censoring, regardless of the correct specification of the propensity score model. 

\subsection{IPTW Cox models}
We will first consider the estimator formed by weighting the partial likelihood estimating equation corresponding to the model in \eqref{cox}, often referred to as a propensity score weighted Cox model. To be explicit, in this Section, the estimator for the causal conditional log hazard ratio we are discussing is the $\beta$ within the vector $\boldsymbol{\theta}=\{\beta, \boldsymbol{\gamma}\}$ that solves:
\begin{eqnarray}
H(\boldsymbol{\theta})=\sum_{l=1}^n \left[\begin{array}{c}\Delta_l W(X_l,\boldsymbol{Z}_{l};\widehat{\boldsymbol{\alpha}})\left\{\begin{array}{c} X_l -\frac{\sum_{j \in R(T_l)}W(X_j,\boldsymbol{Z}_{j};\widehat{\boldsymbol{\alpha}})X_j\exp(\beta X_j+m(\boldsymbol{Z}_j;\boldsymbol{\gamma}))}{\sum_{j \in R(T_l)}W(X_j,\boldsymbol{Z}_{j};\widehat{\boldsymbol{\alpha}})\exp\{\beta X_j+m(\boldsymbol{Z}_j;\boldsymbol{\gamma})\}}\\
m'(\boldsymbol{Z}_l; \boldsymbol{\gamma})-\frac{\sum_{j \in R(T_l)}W(X_j,\boldsymbol{Z}_{j};\widehat{\boldsymbol{\alpha}})m'(\boldsymbol{Z}_j; \boldsymbol{\gamma})\exp(\beta X_j+m(\boldsymbol{Z}_j;\boldsymbol{\gamma}))}{\sum_{j \in R(T_l)}W(X_j,\boldsymbol{Z}_{j};\widehat{\boldsymbol{\alpha}})\exp\{\beta X_j+m(\boldsymbol{Z}_j;\boldsymbol{\gamma})\}}\\
\end{array}\right\}
\end{array}\right]=0
\label{IPTWcox}
\end{eqnarray}
where $m'(\boldsymbol{Z}_l; \boldsymbol{\gamma})=\frac{d m(\boldsymbol{Z}_l; \boldsymbol{\gamma})}{d\boldsymbol{\gamma}}$. We call the solutions to (\ref{IPTWcox}) $\widehat{\beta}^{ipw}$ and $\boldsymbol{\widehat{\gamma}}^{ipw}$.

In addition, in this setting, the regression standardization estimator for the causal survival difference at time $t$ is given by: 
 \begin{eqnarray}
\frac{1}{n} \left\{\sum_{i=1}^n \widehat{S}(t|1, \boldsymbol{Z}_i; \widehat{\beta}^{ipw}, \boldsymbol{\widehat{\gamma}}^{ipw})-  \sum_{i=1}^n \widehat{S}(t| 0, \boldsymbol{Z}_i; \widehat{\beta}^{ipw}, \boldsymbol{\widehat{\gamma}}^{ipw})\right\}, \label{stdcox}
\end{eqnarray}
where $\widehat{S}(t| x, \boldsymbol{Z}_i; \widehat{\beta}^{ipw}, \boldsymbol{\widehat{\gamma}}^{ipw}) =\exp - \left[\exp\{\widehat{\beta}^{ipw}x+m(\boldsymbol{Z}_i;\boldsymbol{\widehat{\gamma}}^{ipw})\}\widehat{\Lambda}_0(t) \right] $
and $\widehat{\Lambda}_0(t)$ is the weighted Breslow estimator (using the above weights) of the cumulative baseline hazard. 

Regardless of the type of censoring, or the correct specification of the propensity score model for confounding, if the working outcome model in \eqref{cox} is not correctly specified for censoring and confounding, the estimator $\widehat{\beta}^{ipw}$ from \eqref{IPTWcox} will not generally be consistent for $\psi(t,\boldsymbol{Z}; \boldsymbol{\omega})$. This is because the log hazard ratio is not collapsible. Thus, if the outcome model is not correctly specified for censoring and confounding, the estimator $\widehat{\beta}^{ipw}$ will be conditional on a different set, or a different functional form of the same set, of covariates than $\psi(t;\boldsymbol{Z, \boldsymbol{\omega}})$. 

However, the lack of collapsibility is not the only reason that the IPTW Cox model does not provide a DR estimator of interest, as even after regression standardization, the estimator given in \eqref{stdcox}, will not generally be consistent for $\zeta(t)$ unless the working outcome model is correctly specified for confounding and censoring. Other IPTW non-collapsible estimands obtained via IPTW of adjusted models, such as the IPTW logistic regression model, are DR after regression standardization, as outlined in \citet{gabriel2023inverse}. Thus, it is not only the lack of collapsibility but additionally the form of the estimator itself, such as the non-canonical link GLM models \citep{gabriel2023inverse}, which results in the lack of the DR property even after regression standardization and even for a collapsible estimand.

The only setting where there are guarantees about consistency when the working outcome model is misspecified is under the null, $\psi(t,\boldsymbol{Z}; \boldsymbol{\omega}) = 0$. We now discuss the robustness properties under the null in settings with different censoring scenarios.

\paragraph{Completely independent censoring under the null}
Under completely independent censoring, i.e. $C \independent (T,X, \boldsymbol{Z})$, and under the null $\psi(t,\boldsymbol{Z}; \boldsymbol{\omega}) = 0$, the estimator $\widehat{\beta}^{ipw}$ obtained from solving \eqref{IPTWcox}, based on the working propensity score weights from model \eqref{ipw} that is correctly specified for confounding, and the working outcome model in \eqref{cox}, which is misspecified, will be consistent when $\psi(t,\boldsymbol{Z}; \boldsymbol{\omega})=0$. We prove this in the supporting information. Now consider the estimator of the true causal marginal survival difference at time $t$ given in \eqref{stdcox} in this same setting of completely independent censoring and under the null. Here the estimator in \eqref{stdcox} will be consistent for $\zeta(t)=0$. This follows from the consistency of the estimator $\widehat{\beta}^{ipw}$ under the null in this setting and the consistency of the Breslow estimator.

\paragraph{Covariate dependent censoring type A under the null}
Type A covariate-dependent censoring is the case where the censoring time is independent of the true survival time conditional on both the exposure and covariates, $T \independent C| (X, \boldsymbol{Z})$, but also that censoring $C$ has a hazard as given by \eqref{sludcens} as in \citet{kong1997robust}. This can clearly occur whenever censoring only depends on the exposure or only on covariates but not both. In this setting, using the working outcome model in \eqref{cox} and working propensity score model \eqref{ipw}, the estimator $\widehat{\beta}^{ipw}$ and the standardization estimator \eqref{stdcox} will have the same properties as they would in the completely independent censoring setting. Thus, under the null $\psi(t,\boldsymbol{Z}; \boldsymbol{\omega})=0$, $\widehat{\beta}^{ipw}$ will be consistent for $\psi(t,\boldsymbol{Z};\boldsymbol{\omega})$ and the standardization estimator given in \eqref{stdcox} will be consistent for $\zeta(t)$. Whenever censoring depends on an interaction of between $X$ and $\boldsymbol{Z}$ there are no guarantees of consistency even under the null. 

\paragraph{Covariate dependent censoring type B under the null}
Consider censoring that is only independent of the true survival time conditional on both the exposure and covariates, $T \independent C| (X, \boldsymbol{Z})$, and the hazard for the censoring contains an interaction between $X$ and $\boldsymbol{Z}$. In this setting, if the Cox working outcome model is not correctly specified for censoring and confounding, even if the propensity score model is correctly specified for confounding, the resulting estimator $\widehat{\beta}^{ipw}$ is not guaranteed to be consistent for $\psi(t,\boldsymbol{Z};\boldsymbol{\omega})$, even under the null. 

For the regression standardization estimator given in \eqref{stdcox} for the causal marginal survival difference at time $t$, we are again reliant on the Cox outcome model to be correctly specified for censoring and confounding for it to be consistent for $\zeta(t)$. 

The guarantee of robustness under the null can potentially be regained in this setting if one includes inverse probability of censoring weights (IPCW) in addition to IPTW. However, both the working models that the IPCW and IPTW rely on must be correctly specified for censoring and confounding, respectively, to allow for the outcome model to be misspecified and maintain consistency under the null. Inverse probability of censoring weighting is described in detail in Subsection \ref{DREE}.

\paragraph{Noncorrectable dependent censoring under the null}
When the censoring is dependent on the survival time conditional on $X$ and $\boldsymbol{Z}$, i.e. informative, there are no general guarantees of either $\widehat{\beta}^{ipw}$ or the regression standardization estimator given by \eqref{stdcox} being consistent for their respective estimands $\psi(t; \boldsymbol{Z}, \boldsymbol{\omega})$ and $\zeta(t)$, even under the null. This is because we have no way of accounting for this dependent censoring in the Cox outcome model nor any other outlet (e.g., IPCW) for modeling the censoring mechanism. However, there may be some special cases, even with informative censoring, where there will be consistency under the null. 

\subsection{IPTW Parametric proportional hazards models}
We now consider the estimator formed by weighting the likelihood estimating equation corresponding to the model in \eqref{cox} where the baseline hazard is specified. To be explicit, in this Section the estimator for the causal conditional log hazard ratio we are discussing is the $\beta$ within the vector $\boldsymbol{\aleph}=\{\beta, \boldsymbol{\gamma}, \nu \}$ that solves:
\begin{eqnarray}
H(\boldsymbol{\aleph})=\sum_{i=1}^n W(X_i,\boldsymbol{Z}_{i};\widehat{\boldsymbol{\alpha}})\frac{\partial}{\partial\boldsymbol{\aleph^T}}[\Delta_i\text{log}\left\{p(T_i|X_i,\boldsymbol{Z}_{i};\boldsymbol{\aleph})\right\}+(1-\Delta_i)\text{log}\left\{S(T_i|X_i,\boldsymbol{Z}_{i};\boldsymbol{\aleph})\right\}]=0,
\label{IPTWpar}
\end{eqnarray}
where the density $p$ and survival function $S$ depend on the specific distribution. The solutions for $\boldsymbol{\aleph}$ to this set of estimating equations we call $\hat{\boldsymbol{\aleph}}=\{\widehat{\beta}^{ipw**}$,  $\boldsymbol{\widehat{\gamma}}^{ipw**}$ and $\widehat{\nu}^{ipw**}\}$.

In addition, in this setting, the regression standardization estimator for the survival difference is given by 
\begin{eqnarray}
\frac{1}{n} \left\{\sum_{i=1}^n S(t| 1, \boldsymbol{Z}_i; \hat{\boldsymbol{\aleph}})-  \sum_{i=1}^n S(t| 0, \boldsymbol{Z}_i; \hat{\boldsymbol{\aleph}})\right\} \label{stdparm}
\end{eqnarray}
where 
\begin{eqnarray*}
S(t| x, \boldsymbol{z}_i; \hat{\boldsymbol{\aleph}}) =\exp \left[-\int^{t}_{0}  \lambda_0(u, \widehat{\nu}^{ipw**})\, \exp\{\widehat{\beta}^{ipw**}x+m(\boldsymbol{z}_i;\boldsymbol{\widehat{\gamma}}^{ipw**})\} du  \right].
\end{eqnarray*}
This model form covers most, if not all, commonly used parametric proportional hazards models, including Cox, Weibull, Exponential and flexible parametric proportional hazards \citep{royston2002flexible}, in addition to less common models such as piecewise exponential, and Gompertz-Makeham \citep{andersen2012statistical}.  

These estimators all have the same properties as the IPTW Cox as outlined above. This means, when the working propensity score model in \eqref{ipw} is correctly specified for confounding, but the working outcome model in \eqref{cox} 
is not correctly specified, $\widehat{\beta}^{ipw**}$ is only guaranteed to be consistent for $\psi(t; \boldsymbol{Z},\boldsymbol{\omega})$, and the estimator in \eqref{stdparm} for $\zeta(t)$, under the null in the setting of Type A covariate dependent censoring, a special case of which is completely independent censoring. Additionally, just like for the IPTW Cox estimators, there are no guarantees of consistency away from the null or under different types of censoring even under the null. 

We prove in the supporting information the robustness under the null of the general IPTW proportional hazards models given in \eqref{cox}, which includes the Cox model and parametric proportional hazards model. Additionally we demonstrate the lack of double robustness away from the null and under different forms of censoring in simulations.

\section{Doubly Robust Estimators and Inference} \label{rec}
We consider two simple doubly robust estimators for the survival difference at a given time point and one more complex doubly robust estimator for the survival curve process. We first consider the simple estimators that are doubly robust in the sense that they are consistent if the censoring model is correctly specified for censoring and either the outcome model is correctly specified for confounding or the exposure model is correctly specified for confounding. 

 
\subsection{Simple doubly robust estimators for the survival difference}
Two simple DR estimators were developed in \cite{blanche2023logistic} and \cite{wang2018simple}, respectively. As presented in \cite{blanche2023logistic}, they are most useful under completely independent censoring or censoring mechanisms that can be easily modeled because the censoring model must be correctly specified for them to be consistent regardless of the correct specification of the outcome models. We highlight the method of 
\cite{blanche2023logistic} here as it is potentially the easiest for practitioners to use and understand as it uses standard logistic regression. However, for practitioners familiar with pseudo-observations \cite{wang2018simple} may be the easiest method to implement, particularly under completely independent censoring. We outline the method of \cite{wang2018simple} in Section S1 of the supporting information and provide code for running all suggested methods in Section S2 of the supporting information.

\paragraph{DR binomial regression} \label{DREE}
As presented in \cite{blanche2023logistic}, we model the causal survival probabilities $p\{T(x)>t\}$. We assume there is a way to correctly specify for confounding a model for $p\{T(x)\leq t|Z\}$ as in a logistic regression model
\[
p\{T \leq t | X, \boldsymbol{Z}\} = Q_t(X, \boldsymbol{Z}; \beta^\dagger, \boldsymbol{\gamma}^\dagger) = \frac{\exp\{\beta^\dagger X+h(X,\boldsymbol{Z};\boldsymbol{\gamma}^\dagger)\}}{1 + \exp\{\beta^\dagger X+h(X,\boldsymbol{Z};\boldsymbol{\gamma}^\dagger)\}}. 
\]
Again, the working propensity model is as given in \eqref{ipw}. In addition to these models, we model the censoring distribution as 
$
p\{C > u | X, \boldsymbol{Z}\} = G_c(u, X, \boldsymbol{Z}), 
$
for some specified parameterization $G_c$ of the conditional censoring survivor function. 
We assume an estimator $\widehat{G}_c(u, X, \boldsymbol{Z})$ can be obtained, for example by a Cox model, Aalen's additive hazard model, nonparametric estimators stratified on covariates, or if censoring is assumed independent of covariates, $G_c$ may be estimated with the Kaplan-Meier estimator or any other nonparametric estimator. Let the estimated inverse probability of censoring weights be
\[
\widehat{U}_i(t) = \frac{I_{\widetilde{T_i} \leq t} \Delta_i}{\widehat{G}_c(\widetilde{T}_i, X_i, \boldsymbol{Z}_i)} + \frac{I_{\widetilde{T_i}> t}}{\widehat{G}_c(t, X_i, \boldsymbol{Z}_i)}. \label{IPCW}
\]
Then let $\widehat{\beta}^\dagger, \widehat{\boldsymbol{\gamma}}^\dagger$ denote the parameter estimates resulting from the inverse probability of censoring weighted, but propensity score unweighted, score equations for the logistic model estimated via maximum likelihood among the subset of individuals who either had the event before time $t$ or remained under observation until at least time $t$. Then the doubly robust estimator of the causal event probability (one minus survival) under exposure level $x = 1$ is
\begin{eqnarray}
\frac{1}{n}\sum_{i = 1}^n W(1, \boldsymbol{Z}_i; \widehat{\boldsymbol{\alpha}})\widehat{U}_i(t)\left\{I_{T_i \leq t} - Q_t(1, \boldsymbol{Z}_i, \widehat{\beta}^\dagger, \widehat{\boldsymbol{\gamma}}^\dagger)\right\} + \frac{1}{n}\sum_{i = 1}^n Q_t(1, \boldsymbol{Z}_i, \widehat{\beta}^\dagger, \widehat{\boldsymbol{\gamma}}^\dagger),  \label{PBE}
\end{eqnarray}
where the term $I_{T_i \leq t}$ is defined to be 0 for individuals censored before time $t$, but it is otherwise fully observed. Similarly, we can substitute in $x = 0$ and take the difference between them to obtain estimates of $\zeta(t)$. A standard error estimator is suggested in \citet{blanche2023logistic} and implemented in the R package \texttt{mets}. Note that there are other similar DR estimators that are discussed in \citet{blanche2023logistic}.  

When censoring is not completely independent or easily modeled, one may wish to use a more complex method that also allows for the censoring model to be misspecified if the outcome model is correctly specified for confounding and censoring. Additionally, this method may be preferred when the full survival curves are of interest rather than just a single point in time. 

\subsection{Doubly robust estimator for the survival curve}
Following \citet{bai2013doubly} and \citet{sjolander2017doubly}, consider the following estimating equation for $S_x(t) = p\{T(x) > t\}$:
\begin{eqnarray*}
\sum_{i = 1}^n \left\{S_x(t) - \frac{I_{X_i = x} I_{\widetilde{T}_i > t}}{\overline{g}(\boldsymbol{Z}_i, \boldsymbol{\alpha})G_c(t, X_i, \boldsymbol{Z}_i)} - \frac{I_{X_i = x} - \overline{g}(\boldsymbol{Z}_i, \boldsymbol{\alpha})}{\overline{g}(\boldsymbol{Z}_i, \boldsymbol{\alpha})} H(t, \boldsymbol{Z}_i, X_i = x) - \right. \\ 
\left. \frac{I_{X_i = x} - \overline{g}(\boldsymbol{Z}_i, \boldsymbol{\alpha})}{\overline{g}(\boldsymbol{Z}_i, \boldsymbol{\alpha})} H(t, \boldsymbol{Z}_i, X_i = x) \int_0^t\frac{d\, M_c(u, \boldsymbol{Z}_i, X_i, \widetilde{T}_i, \Delta_i)}{G_c(u, X_i, \boldsymbol{Z}_i)H(u, \boldsymbol{Z}_i, X_i)}\right\} = 0,
\end{eqnarray*}
where $\overline{g}(\boldsymbol{Z}_i, \boldsymbol{\alpha}) = g(\boldsymbol{Z}_i, \boldsymbol{\alpha})^x \{1 - g(\boldsymbol{Z}_i, \boldsymbol{\alpha})\}^{(1-x)}$, $H(t, \boldsymbol{Z}, X)$ is $p(T > t | \boldsymbol{Z}, X)$, and $M_c(t, \boldsymbol{Z}, X, \widetilde{T}, \Delta)$ is the martingale for the censoring distribution, i.e., the counting process minus the true intensity process for censoring. Please see \citet{bai2013doubly} for precise definitions or the Supplementary Materials just after Equation (2) for a general definition of a martingale. The above is an unbiased estimating equation for $S_x(t)$ if either $H(t, \boldsymbol{Z}, X)$ is correctly specified for both censoring and confounding, i.e., is a correctly specified model for $p\{T(x) > t | \boldsymbol{Z}, X\}$ or both $G_c(u, X, \boldsymbol{Z})$ and $g(\boldsymbol{Z}, \boldsymbol{\alpha})$ are correctly specified for censoring and confounding, respectively. To obtain estimates of the difference in survival probabilities, one must specify models for the unknown functions $g$, $G_c$, and $H$, get estimates of those, plug them into the estimating equations, and solve for $S_x(t)$ under $x \in \{0, 1\}$. In the simulation study we used parametric Weibull survival models as implemented in \texttt{survreg} for the outcome and the censoring distributions, and logistic regression for the propensity score model. \citet{bai2013doubly} provides an expression for a variance estimator that accounts for the uncertainty due to the estimation of the propensity score $g$ and the censoring distribution $G_c$. 

We summarize the double robustness or lack of double robustness of all methods discussed above in Table \ref{summarytab1}. 

\begin{table}[]
    \centering
      \caption{Summary of double robustness of methods reviewed. IPTW=inverse probability of treatment weighted, PH=proportional hazards, DR=doubly-robust.}
    \label{summarytab1}
    \begin{tabular}{p{2cm}p{4cm}p{2cm}p{2cm}p{3cm}}
    \hline
    Estimand&Method & Setting & Censoring type & DR\\
    \hline\\ 
    \hline\\
       Conditional HR & IPTW Cox or IPTW  PH  &  Null & Type A & Yes\\
           \hline 
           Survival difference & IPTW Cox or IPTW PH + Standardization &  Null & Type A & Yes\\
               \hline 
            Conditional HR  & IPTW Cox or IPTW PH  &  Not null & Any type & Not generally\\
                \hline 
                Survival difference & IPTW Cox or IPTW PH  &  Not null & Any type & Not generally\\
              \hline 
              Survival difference&Binomial regression or pseudo-observations - Correctly specified for censoring & Any & Any correctable & Yes\\
                  \hline 
              Survival difference & Binomial regression or pseudo-observations - Misspecified for censoring & Any & Any correctable & Not generally \\
                  \hline 
              Survival curve & DR survival curve & Any & Any correctable& Yes \\             
              \hline \\
    \end{tabular}
  
\end{table}

\section{Simulations} \label{sims}
The data generation and analysis methods for the simulation study are given in the supporting information Section S2.

\subsection{Results}

\begin{table}[ht]
\footnotesize
\caption{Entries are the mean bias (sd) and type I error (T1E) of the conditional log hazard ratio over 1000 simulation replicates with a sample size of 2000 for each replicate. Null = no effect of $X$ on the outcome, non Null = true causal conditional log hazard ratio equals $0.69$. \label{simtab:hr} }
\centering
\begin{tabular}{l|rr|rr|rr}
  \hline
Model specification & \multicolumn{2}{c|}{IPTW Cox} & \multicolumn{2}{c|}{IPTW flexible parametric} & \multicolumn{2}{c}{IPTW Weibull} \\ 
 \hline
  & \multicolumn{6}{c}{Independent censoring, true value = 0} \\
  & bias (sd) & T1E & bias (sd) & T1E & bias (sd) & T1E  \\
  \hline 
  right weights, wrong outcome & -0.004 (0.048) & 0.05 & -0.004 (0.051) & 0.05 & -0.004 (0.051) & 0.06 \\ 
  wrong weights, right outcome & -0.002 (0.053) & 0.03 & -0.003 (0.053) & 0.03 & -0.003 (0.053) & 0.03 \\ 
  wrong weights, wrong outcome & -0.212 (0.055) & 0.97 & -0.218 (0.057) & 0.97 & -0.218 (0.057) & 0.97 \\ 
   \hline
    & \multicolumn{6}{c}{Independent censoring, non Null, true value = 0.69} \\
     & \multicolumn{2}{c|}{bias (sd)} & \multicolumn{2}{c|}{bias (sd)} & \multicolumn{2}{c}{bias (sd)}  \\
    \hline
  right weights, wrong outcome & \multicolumn{2}{c|}{-0.203 (0.047)} & \multicolumn{2}{c|}{-0.191 (0.050)} & \multicolumn{2}{c|}{-0.190 (0.050)} \\ 
  wrong weights, right outcome & \multicolumn{2}{c|}{0.003 (0.056)} & \multicolumn{2}{c|}{0.003 (0.056)} & \multicolumn{2}{c|}{0.004 (0.055)} \\ 
  wrong weights, wrong outcome & \multicolumn{2}{c|}{-0.410 (0.057)} & \multicolumn{2}{c|}{-0.403 (0.059)} & \multicolumn{2}{c|}{-0.403 (0.059)} \\ 
    \hline
    & \multicolumn{6}{c}{Covariate-dependent censoring type A, true value = 0} \\
  & bias (sd) & T1E & bias (sd) & T1E & bias (sd) & T1E  \\
  \hline 
  right outcome & -0.000 (0.060) & 0.05 & -0.000 (0.060) & 0.05 & -0.000 (0.060) & 0.04 \\ 
  wrong outcome & -0.000 (0.061) & 0.06 & 0.006 (0.063) & 0.06 & 0.006 (0.063) & 0.06 \\ 
   \hline
    & \multicolumn{6}{c}{Covariate-dependent censoring type A, non Null, true value = 0.69} \\
     & \multicolumn{2}{c|}{bias (sd)} & \multicolumn{2}{c|}{bias (sd)} & \multicolumn{2}{c}{bias (sd)}  \\
    \hline
 right outcome & \multicolumn{2}{c|}{0.004 (0.057)} & \multicolumn{2}{c|}{0.004 (0.056)} & \multicolumn{2}{c|}{0.004 (0.056)} \\ 
  wrong outcome & \multicolumn{2}{c|}{-0.130 (0.057)} & \multicolumn{2}{c|}{-0.114 (0.058)} & \multicolumn{2}{c|}{-0.114 (0.058)} \\ 
   \hline
    & \multicolumn{6}{c}{Covariate-dependent censoring type B, true value = 0} \\
  & bias (sd) & T1E & bias (sd) & T1E & bias (sd) & T1E  \\
  \hline 
  right outcome & 0.002 (0.060) & 0.05 & 0.002 (0.059) & 0.05 & 0.002 (0.059) & 0.05 \\ 
  wrong outcome & 0.081 (0.061) & 0.28 & 0.100 (0.061) & 0.39 & 0.100 (0.061) & 0.39 \\ 
   \hline
    & \multicolumn{6}{c}{Covariate-dependent censoring type B, non Null, true value = 0.69} \\
     & \multicolumn{2}{c|}{bias (sd)} & \multicolumn{2}{c|}{bias (sd)} & \multicolumn{2}{c}{bias (sd)}  \\
    \hline
 right outcome & \multicolumn{2}{c|}{0.005 (0.059)} & \multicolumn{2}{c|}{0.006 (0.058)} & \multicolumn{2}{c|}{0.006 (0.058)} \\ 
  wrong outcome & \multicolumn{2}{c|}{-0.080 (0.059)} & \multicolumn{2}{c|}{-0.051 (0.059)} & \multicolumn{2}{c|}{-0.051 (0.059)} \\  
   \hline
   & \multicolumn{6}{c}{Outcome-dependent censoring, true value = 0} \\
  & bias (sd) & T1E & bias (sd) & T1E & bias (sd) & T1E  \\
  \hline 
  right weights, wrong outcome & -0.034 (0.081) & 0.09 & -0.030 (0.080) & 0.08 & -0.030 (0.080) & 0.08 \\ 
   wrong weights, wrong outcome & -0.508 (0.071) & 1.00 & -0.501 (0.069) & 1.00 & -0.501 (0.069) & 1.00 \\ 
   \hline
    & \multicolumn{6}{c}{Outcome-dependent censoring, non Null, true value = 0.69} \\
     & \multicolumn{2}{c|}{bias (sd)} & \multicolumn{2}{c|}{bias (sd)} & \multicolumn{2}{c}{bias (sd)}  \\
    \hline
 right weights, wrong outcome & \multicolumn{2}{c|}{-0.324 (0.067)} & \multicolumn{2}{c|}{-0.321 (0.067)} & \multicolumn{2}{c|}{-0.322 (0.067)} \\ 
  wrong weights, wrong outcome & \multicolumn{2}{c|}{-0.803 (0.065)} & \multicolumn{2}{c|}{-0.801 (0.064)} & \multicolumn{2}{c|}{-0.801 (0.064)} \\ 
   \hline
\end{tabular}
\end{table}

\begin{table}[ht]
\footnotesize
\caption{Entries are the mean bias or relative bias in the non-null cases (sd), and type I error (TIE) of the survival difference over 1000 simulation replicates with a sample size of 2000 for each replicate. Null = no effect of $X$ on the outcome, nonNull = true survival difference nonzero. Blank entries are places where a method cannot be used, for example the binomial regression estimator without IPCW model \label{simtab:sdiff} }
\centering
\begin{tabular}{l|rr|rr|rr|rr}
  \hline
Model specification & \multicolumn{2}{c|}{standardized IPTW Cox} & \multicolumn{2}{c|}{DR Bin. regression} & \multicolumn{2}{c|}{DR survival} & \multicolumn{2}{c}{DR pseudo-obs.} \\ 
 \hline
  & \multicolumn{8}{c}{Independent censoring, true value = 0} \\
  & bias (sd) & T1E & bias (sd) & T1E & bias (sd) & T1E & bias(sd) & T1E \\
  \hline 
  right weights, wrong outcome & 0.001 (0.015) & 0.05 & 0.001 (0.022) & 0.04 & 0.001 (0.020) & 0.05 & 0.001 (0.020) & 0.04 \\ 
  wrong weights, right outcome & 0.001 (0.011) & 0.05 & 0.001 (0.021) & 0.04 & 0.000 (0.018) & 0.04 & 0.001 (0.019) & 0.05 \\ 
  wrong weights, wrong outcome & 0.063 (0.016) & 0.99 & 0.063 (0.023) & 0.78 & 0.063 (0.021) & 0.83 & 0.063 (0.022) & 0.83 \\ 
   \hline
    & \multicolumn{8}{c}{Independent censoring, non Null, true value = -0.09} \\
     & \multicolumn{2}{c|}{relative bias \% (sd)} & \multicolumn{2}{c|}{relative bias \%  (sd)} & \multicolumn{2}{c|}{relative bias \%  (sd)} & \multicolumn{2}{c}{bias (sd)}  \\
    \hline
  right weights, wrong outcome & \multicolumn{2}{c}{-21.2\% (0.008)}  &\multicolumn{2}{c}{ -0.1\% (0.0017)} &\multicolumn{2}{c}{ 0.0\% (0.017)} &\multicolumn{2}{c}{ -0.1\% (0.017)}  \\ 
  wrong weights, right outcome & \multicolumn{2}{c}{-0.1\% (0.008) } & \multicolumn{2}{c}{-0.5\% (0.016) }& \multicolumn{2}{c}{-0.3\% (0.015) }& \multicolumn{2}{c}{-0.3\% (0.020) } \\ 
  wrong weights, wrong outcome & \multicolumn{2}{c}{-82.3\% (0.012) } & \multicolumn{2}{c}{-77.1\% (0.019) }& \multicolumn{2}{c}{-77.1\% (0.019) }& \multicolumn{2}{c}{-77.1\% (0.019) } \\ 
     \hline
     & \multicolumn{8}{c}{Covariate-dependent censoring, type A, true value = 0} \\
  & bias (sd) & T1E & bias (sd) & T1E & bias (sd) & T1E & bias (sd) & T1E \\
  \hline 
 right outcome & 0.000 (0.016) & 0.05 &  &  &  &  &  &  \\ 
  wrong outcome & 0.000 (0.019) & 0.05 &  &  &  &  &  &  \\ 
  right outcome, wrong cens &  &  & 0.001 (0.023) & 0.04 & 0.001 (0.021) & 0.05 & 0.001 (0.022) & 0.06 \\ 
  wrong both &  &  & 0.001 (0.024) & 0.04 & 0.001 (0.024) & 0.05 & 0.000 (0.024) & 0.05 \\ 
  wrong outcome, right cens &  &  & 0.001 (0.024) & 0.04 & 0.000 (0.023) & 0.05 & 0.000 (0.023) & 0.05 \\
   \hline
    & \multicolumn{8}{c}{Covariate-dependent censoring, type A, true value = -0.19} \\
     & \multicolumn{2}{c|}{relative bias \%  (sd)} & \multicolumn{2}{c|}{relative bias \%  (sd)} & \multicolumn{2}{c|}{relative bias \%  (sd)} & \multicolumn{2}{c}{bias (sd)}  \\
    \hline
 right outcome & \multicolumn{2}{c}{-0.1\% (0.021) } &  &  &  &  &  &  \\ 
  wrong outcome & \multicolumn{2}{c}{6.0\% (0.026)} &  &  &  &  &  &  \\ 
  right outcome, wrong cens &  &  & \multicolumn{2}{c}{-0.7\% (0.037) } & \multicolumn{2}{c}{-0.0\% (0.034) } & \multicolumn{2}{c}{-0.7\% (0.036)}  \\ 
  wrong both &  &  & \multicolumn{2}{c}{-0.5\% (0.038) } & \multicolumn{2}{c}{-1.0\% (0.038) } & \multicolumn{2}{c}{-1.5\% (0.039)}  \\ 
  wrong outcome, right cens &  &  & \multicolumn{2}{c}{-0.2\% (0.040) } & \multicolumn{2}{c}{0.1\% (0.040) } & \multicolumn{2}{c}{-0.1\% (0.040)}  \\ 
   \hline
    & \multicolumn{8}{c}{Covariate-dependent censoring, type B, true value = 0} \\
  & bias (sd) & T1E & bias (sd) & T1E & bias (sd) & T1E & bias (sd) & T1E \\
  \hline 
 right outcome & -0.001 (0.015) & 0.05 &  &  &  &  &  &  \\ 
  wrong outcome & -0.027 (0.019) & 0.35 &  &  &  &  &  &  \\ 
  right outcome, wrong cens &  &  & -0.016 (0.023) & 0.10 & -0.001 (0.022) & 0.05 & -0.012 (0.022) & 0.09 \\ 
  wrong both &  &  & -0.017 (0.025) & 0.09 & -0.020 (0.024) & 0.13 & -0.018 (0.024) & 0.12 \\ 
  wrong outcome, right cens &  &  & -0.001 (0.024) & 0.04 & -0.001 (0.024) & 0.06 & -0.002 (0.024) & 0.04 \\ 
   \hline
    & \multicolumn{8}{c}{Covariate-dependent censoring, type B, true value = -0.19} \\
     & \multicolumn{2}{c|}{relative bias \%  (sd)} & \multicolumn{2}{c|}{relative bias \%  (sd)} & \multicolumn{2}{c|}{relative bias \%  (sd)} & \multicolumn{2}{c}{relative bias \%  (sd)}  \\
    \hline
 right outcome & \multicolumn{2}{c}{0.4\% (0.017) } &  &  &  &  &  &  \\ 
  wrong outcome & \multicolumn{2}{c}{9.4\% (0.020)} &  &  &  &  &  &  \\ 
  right outcome, wrong cens &  &  & \multicolumn{2}{c}{0.3\% (0.023) } & \multicolumn{2}{c}{0.3\% (0.023) } & \multicolumn{2}{c}{0.2\% (0.023)}  \\ 
  wrong both &  &  & \multicolumn{2}{c}{10.0\% (0.025) } & \multicolumn{2}{c}{9.6\% (0.025) } & \multicolumn{2}{c}{9.2\% (0.025)}  \\ 
  wrong outcome, right cens &  &  & \multicolumn{2}{c}{-0.2\% (0.025) } & \multicolumn{2}{c}{0.0\% (0.025) } & \multicolumn{2}{c}{-0.1\% (0.025)}  \\ 
   \hline
   & \multicolumn{8}{c}{Outcome-dependent censoring, true value = 0} \\
  & bias (sd) & T1E & bias (sd) & T1E & bias (sd) & T1E & bias(sd) & T1E \\
  \hline 
   right weights, wrong outcome & 0.010 (0.025) & 0.09 & -0.029 (0.050) & 0.06 & 0.001 (0.056) & 0.11 & 0.012 (0.041) & 0.07 \\ 
  wrong weights, wrong outcome & 0.148 (0.019) & 1.00 & 0.186 (0.034) & 1.00 & 0.134 (0.039) & 0.97 & 0.112 (0.029) & 0.97 \\ 
   \hline
    & \multicolumn{8}{c}{Outcome-dependent censoring, non Null, true value = -0.19} \\
     & \multicolumn{2}{c|}{relative bias \%  (sd)} & \multicolumn{2}{c|}{relative bias \%  (sd)} & \multicolumn{2}{c|}{relative bias \%  (sd)} & \multicolumn{2}{c}{relative bias \%  (sd)}  \\
    \hline
 right weights, wrong outcome  & \multicolumn{2}{c}{8.9\% (0.019) } & \multicolumn{2}{c}{104\% (0.047)} &\multicolumn{2}{c}{ -1.7\% (0.039)} &\multicolumn{2}{c}{ -16.4\% (0.037)}  \\ 
  wrong weights, wrong outcome & \multicolumn{2}{c}{-132.8\% (0.017) } & \multicolumn{2}{c}{-108\% (0.034)} &\multicolumn{2}{c}{ -133\% (0.026)} &\multicolumn{2}{c}{ -132\% (0.029)}  \\ 
   \hline
\end{tabular}
\end{table}

As seen in Table \ref{simtab:hr}, our simulation confirms that the proportional hazards regression models require the outcome model to be correctly specified to consistently estimate the hazard ratio for exposure when it is truly different from 1. In Table \ref{simtab:sdiff}, we see that the methods for directly estimating the difference in survival probabilities are doubly robust when censoring is completely independent or can be modeled. Additionally, the doubly robust survival curve method can be seen to be robust against outcome model misspecification when both  censoring and confounding can be correctly modeled. In contrast, the standardized Cox and Weibull models show small amounts of bias for estimating the survival difference when the outcome model is misspecified. Additionally, one can see that the standard error estimation methods used for the three doubly robust estimation methods control the type one error rate when they are consistent. Table S1 of the supporting information demonstrates the bias of the standard error estimating procedures for each method in Table \ref{simtab:sdiff}. 

To illustrate the potential for weights to increase the variance of the estimators, we ran simulations with a true log hazard ratio of $\log(2)$ under independent censoring with a sample size of 80. The empirical standard error of the estimated log hazard ratio was 0.337 for the unweighted but correctly specified Cox model over 10000 replicates. Comparing that to a correctly specified Cox model with incorrectly specified weights by forcing the inclusion of an extra variable that is independent distributed $t$ with 2 degrees of freedom to the propensity score model, giving weights that range from about 1 to 10, we observe a relative increase in the empirical standard error of 15.7\%. When comparing a misspecified outcome Cox model with misspecified weights to the unweighted misspecified Cox model in a similar way, we observe a more extreme relative difference in the empirical standard error of 38.6\%. Although both estimators are biased, misspecified weights increase the bias by 19.2\%. 

\section{Real Data Example} \label{Real}

We illustrate all methods in Table \ref{simtab:sdiff} using the \texttt{rotterdam} dataset included in the \texttt{survival} R package \citet{royston2013external}. Briefly, the Rotterdam tumor bank includes 2982 primary breast cancers patients. We consider the outcome of recurrence or death at times 2.5, 5, and 7.5 years following primary resection surgery. Our exposure of interest is a binary indicator of receiving chemotherapy. We adjust for year of surgery, age of surgery, menopause, tumor size, grade, number of positive lymph nodes, progesterone receptors, estrogen receptors. The propensity score was estimated using logistic regression for each method containing all baseline variables in the outcome model. Censoring was treated as completely independent, and IPCW were based on the Kaplan Meier estimator for the DR binomial regression and DR survival curve methods.

The results are shown in Table \ref{example}; code to reproduce the results is included in the supporting information Section S4. As one can see in Table \ref{example}, the DR survival curve method fails to reject the null, while all other methods exclude the null from the 95\% CI over the three time points. The standard error estimates for the DR survival method are substantially higher than the other methods. Hence we also used the nonparametric bootstrap for standard error estimation with that method for comparison. 

In testing for the proportional hazard assumptions in the Cox outcome model using Schoenfeld residuals, we reject with a p-value of 0.003. Thus, it is likely that the findings of the DR binomial regression and the DR pseudo observation methods are more robust, as they do not enforce proportional hazards. It is of note that although our selected method for the DR survival curve uses a parametric proportional hazards Weibull model, this is not required, and one could use a more flexible outcome model, particularly in settings where proportional hazards is likely to be violated.

\begin{table}[ht]
\caption{Results of the analysis of the Rotterdam data. Bootstrap denotes that the standard error was estimated using the nonparametric bootstrap. est = estimate, se = estimated standard error, lower/upper 95=lower and upper limits of 95\% confidence interval.\label{example}}
\centering
\begin{tabular}{l|rrrr}
  \hline
model & est & se & lower 95 & upper 95 \\ 
  \hline
  & \multicolumn{4}{c}{2.5 years} \\
  \hline
  standardized IPTW Cox -- bootstrap & 0.048 & 0.018 & 0.012 & 0.084 \\ 
  DR survival & 0.047 & 0.069 & -0.087 & 0.182 \\ 
  DR survival -- bootstrap &  & 0.027 & -0.007 & 0.101 \\ 
  DR binomial regression & 0.059 & 0.024 & 0.011 & 0.106 \\ 
  DR pseudo observations -- bootstrap & 0.055 & 0.025 & 0.006 & 0.104 \\ 
   \hline
  & \multicolumn{4}{c}{5 years} \\
  \hline
  standardized IPTW Cox -- bootstrap& 0.067 & 0.026 & 0.016 & 0.118 \\ 
  DR survival & 0.052 & 0.076 & -0.097 & 0.201 \\ 
  DR survival -- bootstrap&  & 0.034 & -0.014 & 0.118 \\ 
  DR binomial regression & 0.068 & 0.028 & 0.013 & 0.124 \\ 
  DR pseudo observations -- bootstrap & 0.070 & 0.029 & 0.012 & 0.128 \\ 
    \hline
  & \multicolumn{4}{c}{7.5 years} \\
  \hline
  standardized IPTW Cox -- bootstrap & 0.073 & 0.029 & 0.017 & 0.129 \\ 
  DR survival & 0.055 & 0.142 & -0.222 & 0.333 \\ 
  DR survival -- bootstrap & 0.055 & 0.034 & -0.012 & 0.123 \\ 
  DR binomial regression & 0.070 & 0.031 & 0.010 & 0.130 \\ 
  DR pseudo observations -- bootstrap & 0.077 & 0.031 & 0.016 & 0.139 \\ 
   \hline
\end{tabular}
\end{table}

\section{Summary} \label{summary}
We wish to be as clear as possible about the following three points. First, all of the above outcome regression methods are consistent for the given estimand of choice if the outcome model is correctly specified for confounding and censoring. Thus, focusing on the correct specification of the outcome model is of the highest importance for the consistency of the estimates when using, for example, an adjusted Cox model for causal inference. Although this holds regardless of the correctness of the propensity score weights, and thus it would seem there is nothing to lose with including them, misspecified weights may cause increased variance or finite sample bias, something we demonstrate in simulations. 

Second, the propensity score weighting of score equations of a partial or full likelihood of an adjusted proportional hazards model does not generally result in a doubly robust estimator for the causal conditional hazard ratio. The corresponding regression standardization estimator using the results of the weighted procedure for the causal marginal survival difference at a particular time point also has no guarantees of consistency if the outcome model is misspecified. We have shown this in simulations for some parametric and semiparametric (Cox) proportional hazards models.

Third, the only situation where including propensity score weights in the modeling procedure  for a proportional hazards model yields a doubly robust estimator without further assumptions is under the null and under a particular form of censoring. We prove the double robustness property in the supporting information. Furthermore, the standard error estimates used must correctly account for the variation in all models fit. Otherwise, although the estimation may be consistent under the null, the type I error rate may be inflated. 

\section{Discussion} \label{dis}
We have shown that away from the null and under certain types of censoring, even under the null, the IPT-weighted proportional hazards survival model can give inconsistent estimates for the causal conditional log hazard ratio whenever the outcome model is misspecified. Although we do not attempt to prove it, we believe it is likely the case that there is no doubly robust estimator for the causal conditional hazard ratio under the standard parameterization of proportional hazards outcome, propensity score, and censoring models; similar to the conditional odds ratio, as shown in Theorem 3 of \citet{tchetgen2010doubly}. Additionally, even after regression standardization to obtain an estimator for the marginal survival difference from an IPTW PH survival outcome model, we do not obtain a doubly robust estimator for the marginal survival difference. This suggests that it is not only the lack of collapsibility that prevents an IPTW PH model from producing a doubly robust estimator. Although it is possible that there exists some causal parameter that an estimator obtained by combining IPTW and the PH survival outcome models we have considered above would be doubly robust for, we are fairly certain that if there is such a parameter, the estimator is not generally doubly robust for this parameter under all types of censoring. 

We only consider the inclusion of the IPTW via weighting the (partial) score equations. Though there are other methods of accounting for confounding, e.g. matching, we are again reasonably confident that any method that mimics a randomized trial is unlikely to result in a DR estimator when combined with an adjusted proportional hazards model because, even in a perfectly randomized trial, a misspecified proportional hazard model results in biased estimates of the conditional hazard ratio. 

We have focused on the causal conditional hazard ratio and the causal marginal survival difference throughout the paper, suggesting some existing methods of doubly robust estimation for the latter. However, there are other estimands, such as the causal marginal hazard ratio and the causal marginal hazard difference, for which doubly robust estimators have been developed. \citet{dukes2019doubly} develop a doubly robust estimator for the conditional hazard difference, which is implemented in R code included with their paper, while \citet{tchetgen2012parametrization} presents a doubly robust estimator for the causal marginal hazard ratio. We note that the estimand studied in \citet{tchetgen2012parametrization} is the causal version of the marginal hazard ratio discussed in Section 2 of this work, i.e., adjusting for or removing the effect of confounding.

It is of note that we have not considered settings with competing risks. In these settings, interpretations of parameters and calculations of survival differences become more complicated. Thus, a different parameter may be a better target of interest. It is unlikely that any ad hoc combination of Cox model(s) and IPTW will result in a doubly robust estimator for the target parameter, particularly when censoring is in any way dependent. However, it is the case that under the null if censoring is of Type A for all competing event times, we still have double robustness for the causal conditional log hazard ratio and the survival difference after regression standardization. 

The trend of applied researchers constructing ad hoc estimators in this manner should stop, as, at least in general, this will not result in a doubly robust estimator. We note that if the goal is to obtain similar estimates to those from the same outcome model in a randomized clinical trial, regardless of correct outcome model specification, correct propensity score weighting will provide this. However, this requires that the propensity score be correct, something that can never be known outside of a randomized trial.

We caution against the blanket conclusion that IPTW Cox or IPTW parametric proportional hazard models maintain a nominal type I error rate even if the propensity or the survival outcome model is correctly specified for confounding. Although this clearly also depends on the censoring mechanism, as we demonstrate, the type I error rate also strongly depends on the correct estimation of the standard error. The inclusion of IPTW into the procedure requires accounting for the uncertainty in those weights in the standard error estimates.

\section*{Acknowledgments}
EEG was partially supported by a grant from Novo Nordisk fonden NNF22OC0076595, and IW by Vetenskapsrådet 2016-00703.

\section*{Supplementary Materials}
Web Appendices, Tables, and code referenced in the abstract and Sections 2, 3, 4, 5, 6, 7, and 8 are available with this paper at the Biometrics website on Wiley Online Library.

\section*{Data Availability}
The data used in the example are publicly available and distributed as part of the R package \texttt{survival}. It can be downloaded from \url{https://cran.r-project.org/package=survival}.

\bibliographystyle{biom}
\bibliography{refs}

\end{document}